\documentclass[12pt, aps, floatfix, prd, final, notitlepage, oneside, onecolumn, nobibnotes, nofootinbib, superscriptaddress, showpacs]{revtex4-1}
\usepackage{amsmath}
\usepackage{graphicx}
\usepackage{esint}
\usepackage[hidelinks]{hyperref}
\usepackage{graphicx}

\newcommand{\R}{\mathcal{R}}
\newcommand{\M}{\mathcal{M}}
\newcommand{\A}{\mathcal{A}}
\newcommand{\Br}{\mathrm{Br}}
\newcommand{\GeV}{\text{GeV}}
\newcommand{\ps}{\text{ps}}

\begin{document}
\title{Exclusive Decays of the Doubly Heavy Baryon $\Xi_{bc}$}

\author{A. V. Luchinsky}
\affiliation{Institute for High Energy Physics, NRC “Kurchatov Institute”, Protvino 142 281, Russia}

\author{A.K. Likhoded}
\affiliation{Institute for High Energy Physics, NRC “Kurchatov Institute”, Protvino 142 281, Russia}

\begin{abstract}
Exclusive decays  of the $\Xi_{bc}^+$ with light mesons production, $\Xi_{bc}^+\to\Xi_{cc}^{++}\R$ are analyzed. In the framework of the factorization model the matrix elements of the considered reactions are written as a product of $\Xi_{bc}\to\Xi_{cc}W$ and $W\to\R$ transition amplitudes. As a result theoretical and experimental investigation of these decays allow us to test the physics of heavy and light quarks' sectors. 
 Presented are the branching fractions of these reactions, as well as the distributions over the invariant mass of the light system and some other kinematical variables. Our calculations show that the probabilities of the considered reactions are high enough, so presented results could help with observation of yet unseen $\Xi_{bc}$ baryon.
\end{abstract}

\maketitle

\section{Introduction}

Doubly heavy baryons (DHB) are extremely interesting objects that allow us to take a fresh look at a problem of hadronization of heavy quarks. In valence approximation these particles are build from two heavy and one light quark $(Q_1 Q_2 q)$ and one of the ways to describe such states is to consider two of the valence quarks (e.g. $Q_1q$) as a heavy diquark in anti-triplet color state.  This leaves us with the  effective heavy-quarkonium like particle $([Q_1q]_{\bar{3}_c}Q_2)$ and allows us to check all theoretical methods created for heavy quarkonia description on a new set of objects.

There are a lot of theoretical works devoted to the spectroscopy of doubly heavy baryons \cite{Gershtein:1998un,Kiselev:2002iy,Narodetskii:2002ks}, their production cross sections \cite{Berezhnoy:1995fy, Braguta:2000dc,Ma:2003zk}, lifetimes \cite{Likhoded:1999yv,Guberina:1999mx,Karliner:2014gca, Likhoded:2019rik}, and branching fractions of some exclusive decays \cite{Onishchenko:2000wf,Albertus:2007xc,Eakins:2012fq, Gutsche:2019iac,Gutsche:2019wgu,Gutsche:2018msz,Gutsche:2017hux, Pan:2020qqo}. A nice theoretical review for this class of particles can also be found in \cite{Kiselev:2001fw}. Up to recent times, however, such an interest was mostly theoretical since no DHB states were observed experimentally. First experimental result was published  by the SELEX collaboration. In the paper \cite{Ocherashvili:2004hi} it was announced that $\Xi_{cc}^{+}$ baryon was observed in $\Xi_{cc}^{+}\to p D^{+}K^{-}$ decay channel. This result was not confirmed, but later LHCb collaboration managed to observe the other doubly charged $\Xi_{cc}^{++}$ baryon in $\Xi_{cc}^{++}\to\Lambda_{c}^{+}K^-\pi^+\pi^+$ and $\Xi_{cc}^{++}\to\Xi_{c}^{+}\pi^{+}$ decay channels \cite{Aaij:2017ueg,Aaij:2018gfl}.

Currently $\Xi_{cc}^{++}$ baryon is the only DHB particle observed experimentally. We are hoping, however, that discovery of some other particles of this family is on the way. As an interesting example we would like to mention DHB states with mixed heavy flavours, e.g. $\Xi_{bc}^{+}=(bcu)$. The cross sections of its production is expected to be comparable with the cross section of $B_c$ meson production, which was already observed in hadronic experiments \cite{Aaij:2019ths,Aaij:2019ldo}. For this reason it seems very interesting to study in more details some of the $\Xi_{bc}$ baryon's decays.

This topic was also widely discussed in the literature. In our paper we would like to consider in more details the processes of light mesons' production in exclusive $\Xi_{bc}$ decays. In our recent paper \cite{Gerasimov:2019jwp} this problem was studied in the framework of the spectral function approach. Such an approach, however, allows one to obtain only the distributions over the invariant mass of light mesons' system and calculate the integrated branching fractions. For comparison with the future experimental data more detailed theoretical predictions (including distributions over other kinematical variables) will be required. This is the topic of our current paper.
As you can see in mentioned above paper, the branching fractions of light mesons' production $\Xi_{bc}^{+}\to\Xi_{cc}^{++}$ transitions are large enough and, in addition, $\Xi_{cc}^{++}$ baryon was observed already in experiment. This is why one could expect that these decays will be observed in the nearest future, so  in our paper we will concentrate specifically on them.

The rest of the paper is organized as follows. In the next section the matrix elements of the considered decays are presented and the parametrization of the used form factors is given. In section \ref{sec:spectr-funct-form} we describe the spectral function formalism,  expressions for light mesons' production vertices, and give predictions for integrated branching fractions  and transferred momentum distributions obtained with different parameterizations of the form-factors. In section \ref{sec:distr-over-other} distributions over some other kinematical variables are presented and discussed. The results of our work are summarized in the last section.

\section{$\Xi_{bc}^{+}\to\Xi_{cc}^{++}$ Matrix Element and the Form Factors}
\label{sec:xi_bc+t-matr-elem}

\begin{figure}
  \centering
  \includegraphics[width=0.9\textwidth]{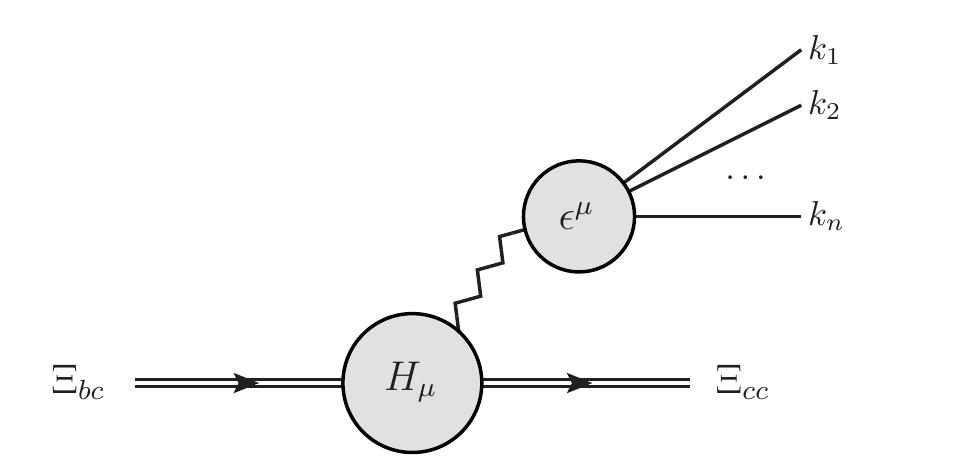}
  \caption{Feynman diagram describing $\Xi_{bc}^{+}\to \Xi_{cc}^{++}\R$ decay}
  \label{diag:main}
\end{figure}

Let us consider an exclusive decay of $\Xi_{bc}^{+}$ baryon with the production of light particles' system $\R$, which could be a semileptonic pair $\ell\nu_{\ell}$, a single $\pi$ meson, $\rho$, or even some larger set of light mesons. At the leading order of the perturbation theory this process is described by the Feynman diagram shown in Fig.~\ref{diag:main}. The corresponding  matrix element can be written in the form
\begin{align}
\M & =  \frac{G_{F}V_{CKM}}{\sqrt{2}}a_{1}H^{\mu}\epsilon_{\mu}^{(\R)},\label{eq:H}
\end{align}
where $\epsilon_{\mu}^{\R}$ is the effective polarization vector of the system $\R$, $H^{\mu}$ is the matrix element of  $\Xi_{bc}^{+}\to\Xi_{cc}^{++}$ transition, and  $a_{1}$ factor describes the effect of soft gluon rescattering \cite{Buchalla:1995vs}. It should be set equal to unity in the case of the semileptonic pair in the final state, and, since we are dealing with $b$-quark decay,
\begin{align}
  a_{1} &= 1.2
\end{align}
in all other cases.

\begin{figure}
  \centering
  \includegraphics[width = 0.9\textwidth]{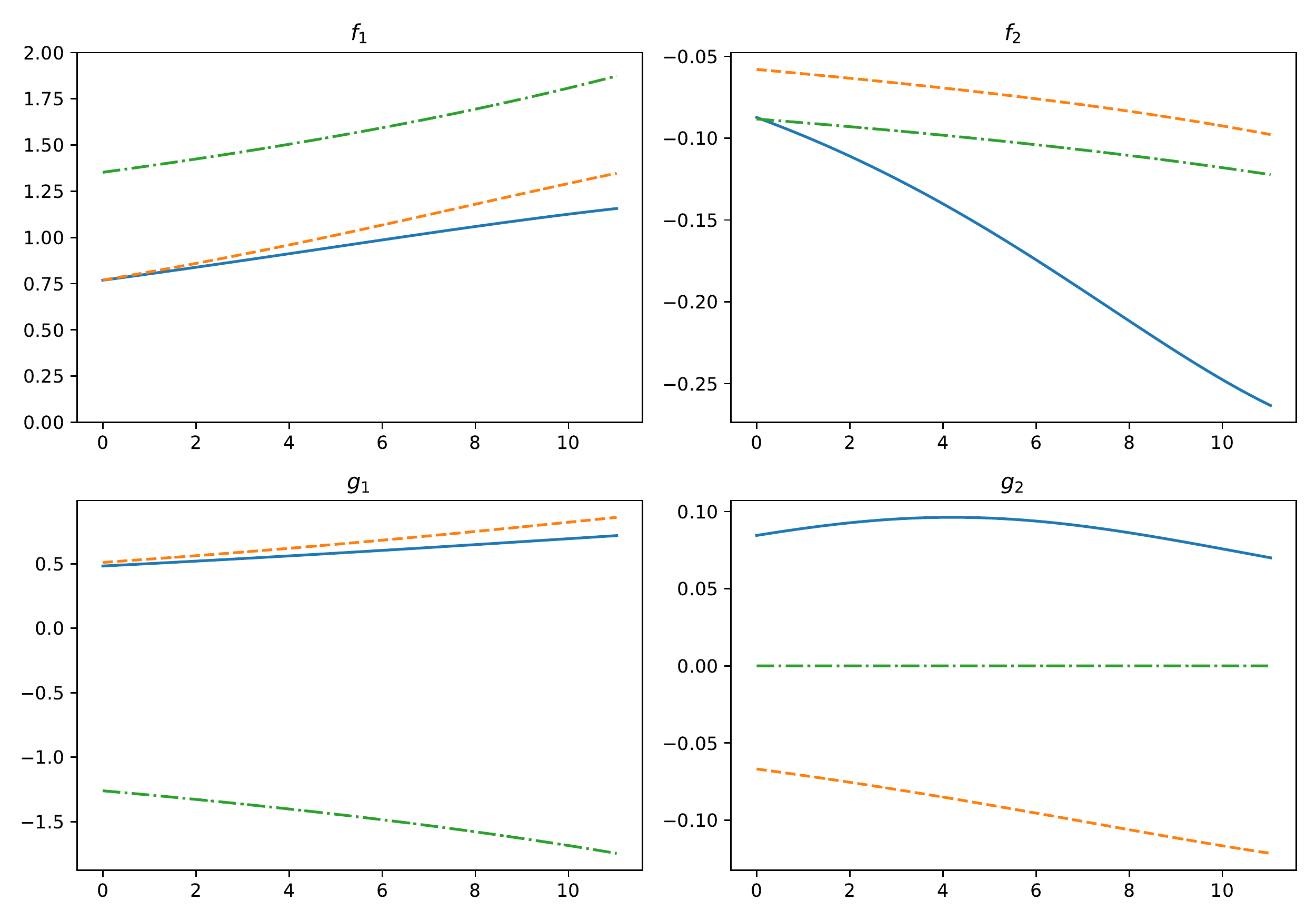}
  \caption{Form factors of the $\Xi_{bc}^{+}\to\Xi_{cc}^{++}$ transition as a functions of squared transferred momentum $q^2$ (in $\GeV^2$). Solid, dashed, and dash-dotted lines correspond to [W\_17], [H\_20], and [O\_00] form factor sets respectively}
  \label{fig:FF}
\end{figure}

The matrix element of the $\Xi_{bc}\to\Xi_{cc}$ transition is  written using the corresponding form factors, and it can be done in several ways. In the works \cite{Wang:2017mqp, Hu:2020mxk}, for example, the parametrization
\begin{align}
H_{\mu} & =  \bar{u}(P_{1})\left[f_1\left(q^2\right)\gamma_{\mu}+i\frac{q^{\mu}}{M_{1}}\sigma_{\mu\nu}f_2\left(q^{2}\right)+\frac{q_{\mu}}{M_{1}}f_3\left(q^{2}\right)\right]u(P_{2})+\nonumber \\
 & \bar{u}(P_{1})\left[g_1\left(q^2\right)\gamma_{\mu}+i\frac{q^{\mu}}{M_{1}}\sigma_{\mu\nu}g_2\left(q^{2}\right)+\frac{q_{\mu}}{M_{1}}g_3\left(q^{2}\right)\right]\gamma_{5}u(P_{2}),\label{eq:VAFF}
\end{align}
was adopted. Here the notation
\begin{align}
  \sigma_{\mu\nu} & = \frac{i}{2}\left(\gamma_{\mu}\gamma_{\nu}-\gamma_{\nu}\gamma_{\mu}\right)
\end{align}
is introduced, $P_{1,2}$ and $q=P_{1}-P_{2}$ are the momenta of final and initial baryons and the transferred momentum respectively, and $M_{1}$ is the mass of the initial particle. In the papers \cite{Onishchenko:2000wf,Likhoded:2009zz}, on the other hand, the vertex of the weak decay is parametrized  as
\begin{align}
  H_{\mu} & =  \bar{u}\left(P_{1}\right)\left[G_{1}^{V}\left(q^{2}\right)\gamma_{\mu}+v_{1\mu}G_{2}^{V}\left(q^{2}\right)+v_{2\mu}G_{3}^{V}\left(q^{2}\right)\right]u\left(P_{2}\right)+\\
 &  \bar{u}\left(P_{1}\right)\gamma_{5}\left[G_{1}^{A}\left(q^{2}\right)\gamma_{\mu}+v_{1\mu}G_{2}^{A}\left(q^{2}\right)+v_{2\mu}G_{3}^{A}\left(q^{2}\right)\right]u\left(P_{2}\right),
\end{align}
where $v_{1,2} = P_{1,2}/M_{1,2}$ are the invariant velocities of the initial and final baryons. The following relations can be used to switch from one parametrization to the other:
\begin{align}
  f_1 & = G_{1}^{V}+\left(M_{1}+M_{2}\right)\left(\frac{G_{2}^{V}}{2M_{1}}+\frac{G_{3}^{V}}{2M_{2}}\right),\\
f_2 & = -\frac{G_{2}^{V}}{2M_{1}}-\frac{G_{3}^{V}}{2M_{2}},\\
f_3 & = -\frac{G_{2}^{V}}{2M_{1}}+\frac{G_{3}^{V}}{2M_{2},},\\
g_1 & = -G_{1}^{A}-\left(M_{1}-M_{2}\right)\left(\frac{G_{1}^{A}}{2M_{1}}+\frac{G_{2}^{A}}{2M_{2}}\right),\\
g_2 & = \frac{G_{2}^{A}}{2M_{1}}+\frac{G_{3}^{A}}{2M_{2}},\\
g_3 & = \frac{G_{2}^{A}}{2M_{1}}-\frac{G_{3}^{A}}{2M_{2}}.
\end{align}
In the following form factors presented in papers \cite{Onishchenko:2000wf}, \cite{Wang:2017mqp}, and \cite{Hu:2020mxk} will be used and we will mark them as [On\_00], [W\_17], and [H\_20] respectively. The parametrization (\ref{eq:VAFF}) will be used for all these form factors' sets. Due to vector and partial current conservation the contributions of $f_3(q^2)$, $g_3(q^2)$ form factors are negligible, while the $q^{2}$ dependence of all others is shown in figure \ref{fig:FF}. All these form factors can be written approximately in the form
\begin{align}
  F(q^{2}) = F(0)\left[
  1 + \alpha_{1}\frac{q^{2}}{M_{1}^{2}} + \alpha_{2}\left(\frac{q^{2}}{M_{1}^{2}}\right)^{2},
  \right]
\end{align}
where $F(0)$, $\alpha_{1,2}$ parameters are given in table \ref{tab:FF}.

\begin{table}
\begin{tabular}{c|ccc|ccc|ccc|}\hline
  & \multicolumn{3}{c}{On\_00} & \multicolumn{3}{c}{W\_17} &  \multicolumn{3}{c|}{H\_20} \\
\hline
 & $F(0)$ & $\alpha_1$ & $\alpha_2$  & $F(0)$ & $\alpha_1$ & $\alpha_2$  & $F(0)$ & $\alpha_1$ & $\alpha_2$ \\ 
\hline  $f_1$ & $1.36$ & $1.12$ & $2.36$  & $0.76$ & $2.40$ & $-0.57$  & $0.77$ & $2.94$ & $1.84$  \\
   $f_2$ & $-0.09$ & $1.12$ & $2.36$  & $-0.08$ & $7.37$ & $9.30$  & $-0.06$ & $1.84$ & $4.73$  \\ 
  $g_1$ & $-1.26$ & $1.12$ & $2.36$  & $0.48$ & $1.88$ & $1.16$  & $0.51$ & $2.38$ & $2.75$  \\ 
  $g_2$ & $-0.00$ & $1.12$ & $2.36$  & $0.08$ & $2.84$ & $-16.21$  & $-0.07$ & $3.33$ & $1.62$  \\
  \hline \end{tabular}

\caption{Form factors' parameters in the case of  $\Xi_{bc}^{+}\to\Xi_{cc}^{++}$ transition\label{tab:FF}}
\end{table}

\section{Spectral Functions Formalism and $q^2$ Distributions}
\label{sec:spectr-funct-form}

To describe in detail all kinematics of decay it is necessary to use various numerical methods. If we are interested in $q^{2}$ distributions or the values of the integrated branching fractions only, it is possible to obtain the analytical expressions with the help of the spectral function formalism. This method was successfully used, for example, for description of $\tau$-lepton \cite{Kuhn:2006nw,Kuhn:1990ad} or $B_c$ meson \cite{Luchinsky:2012rk, Likhoded:2013iua, Luchinsky:2013yla,Aaij:2016rks,Aaij:2013gxa} decays.

In the framework of this approach the transferred momentum distribution is written in the form
\begin{align}
  \frac{d\Gamma}{dq^{2}} &= H_{T}^{2} \rho_{T}^{(\R)}  + H_{L}^{2} \rho_{L}^{(\R)},
\end{align}
where transverse and longitudinal squared matrix elements $H_{T, L}^2$ are equal to
\begin{align}
H_T^2&=H^{\mu}H^{\nu*}(q_{\mu}q_{\nu}-q^2g_{\mu \nu})=
\frac{1}{2M_1^2}\Big[f_1^2 M_1^2 (-2 q^4+2 q^2 M_{-}^2-q^2 M_{+}^2+M_{-}^2 M_{+}^2)+
\nonumber\\ &
12 f_1 f_2 q^2 M_{+} (q^2-M_{-}^2) M_1-4 f_2^2 q^2 (q^4-q^2 M_{-}^2+2 q^2 M_{+}^2-2 M_{-}^2 M_{+}^2)+
\nonumber\\ &
g_1^2 (M_{-}+M_{+})^2 (-2 q^4-q^2 M_{-}^2+2 q^2 M_{+}^2+M_{-}^2 M_{+}^2)+12 g_1 g_2 q^2 M_{-} (M_{+}^2-q^2) (M_{-}+M_{+})-
\nonumber\\ &
4 g_2^2 q^2 (q^4+2 q^2 M_{-}^2-q^2 M_{+}^2-2 M_{-}^2 M_{+}^2)\Big] \nonumber
\\ 
H_L^2&=H^{\mu}H^{\nu*}q_{\mu}q_{\nu}= 2 \bigg[ f_1^2 M_{-}^2
       \left(M_{+}^2-\text{q}^2\right)+g_1^2 M_{+}^2 \left(M_{-}-\text{q}^2\right) \bigg],\\
  M_{\pm} &= M_{1} \pm M_{2},
\end{align}
while the spectral functions $\rho_{L, T}(q^{2})$ are defined by the expression
\begin{align}
  \frac{1}{2\pi}\int \delta^{4}\left(q-\sum k_{i}\right) \prod\frac{d^{3}k_{i}}{2e_{i}(2\pi)^{3}}\epsilon^{(\R)}_{\mu}\epsilon^{*(\R)}_{\nu}
  &= \rho_{L}^{(\R)}q_{\mu}q_{\nu} + \rho_{T}^{(\R)}\left(q_{\mu}q_{\nu} - q^{2}g_{\mu\nu}\right).
\end{align}
The explicit form of the spectral functions depends on the final state $\R$.

In some simple cases it is easy to obtain the analytical expressions for these spectral functions. If we are considering the production of the single $\pi$ meson, for example, the effective polarization vector in (\ref{eq:H}) is equal to
\begin{align}
  \epsilon_{\mu}^{(\pi)} &= f_{\pi}q_{\mu},
\end{align}
so we have
\begin{align}
  \rho_{T}^{(\pi)} &=0, \qquad \rho_{L}^{(\pi)}  = f_{\pi}^{2}\delta\left(q^{2}-m_{\pi}^{2}\right). 
\end{align}
In the case of single $\rho$ meson production the expressions are
\begin{align}
  \epsilon_{\mu}^{(\rho)} &= f_{\rho} m_\rho \epsilon_{\mu}, \qquad
  \rho_{L}^{(\rho{\pi})} =0, \qquad
  \rho_{T}^{(\rho)}  = f_{\rho}^{2}\delta\left(q^{2}-m_{\rho}^{2}\right).   
\end{align}
For the semileptonic decays we have
\begin{align}
  \epsilon_{\mu}^{(\ell\nu)} &= \bar{u}(k_{1})\gamma_{\mu}(1-\gamma_{5})v(k_{2}), \qquad
  \rho_{L}^{(\ell\nu)} =0, \qquad
  \rho_{T}^{(\ell\nu)}  = \frac{1}{6\pi^{2}},   
\end{align}
where the mass of the final lepton $\ell$ is neglected.

In more interesting and complicated cases it is not possible to obtain the analytical expressions for the spectral functions, so we should use some model approaches.  Available experimental data, for example, the processes of the light mesons' production in exclusive $\tau$ lepton decays can also be very helpful. For these processes the $q^{2}$ distributions is equal to
\begin{align}
  \frac{d\Gamma\left(\tau\to\nu_{\tau}\R\right)}{dq^{2}} &=
                                                           \frac{G_{F}^{2}}{16\pi} \frac{\left(m_{\tau}^{2}-q^{2}\right)^{2}}{m_{\tau}^{3}}
                                                           (m_{\tau}^{2}+2q^{2}) \rho_{T}^{(\R)}(q^{2}).
\end{align}
From the analysis of the experimental distribution over this variable one can get information both about the form of the transversal spectral function and its normalization.

\begin{figure}
\includegraphics[width=0.45\textwidth]{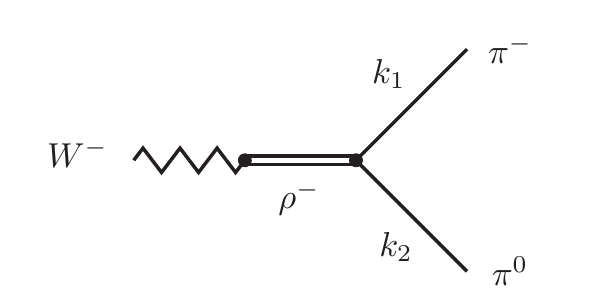}\includegraphics[width=0.45\textwidth]{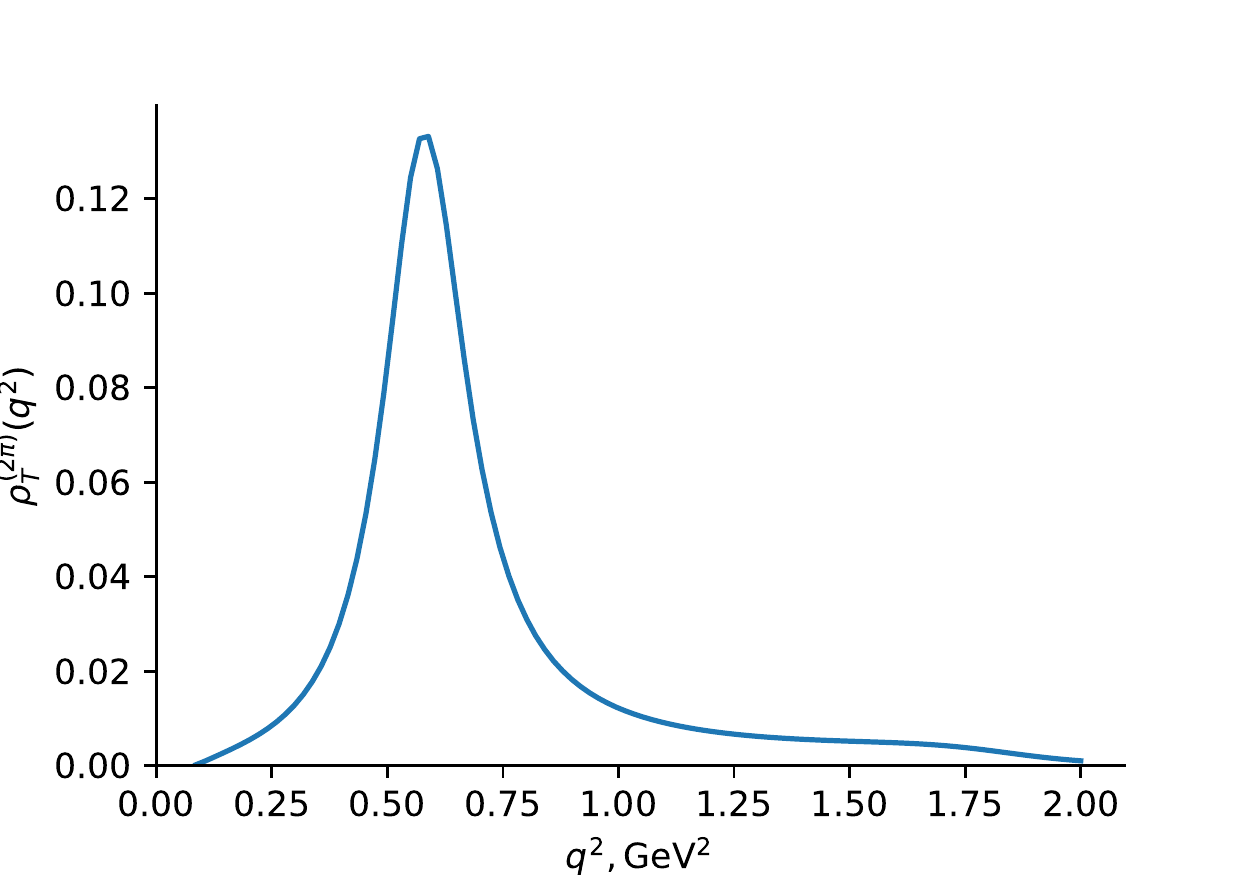}
\caption{Feynman diagram for the  $W\to2\pi$ transition and the corresponding spectral function\label{fig:rt2pi}}
\end{figure}

Let us first consider the exclusive production of $\pi^{-}\pi^{0}$ pair. In the resonance approach this process is described by the diagram shown in figure \ref{fig:rt2pi}. As you can see from this diagram, we are saturating the process under consideration by the contributions of the $\rho$ meson and its excitations. Up to overall normalization the analytical form of the corresponding amplitude can be guessed from the quantum numbers of the participating in the reaction particles:
\begin{align}
  \epsilon_{\mu}^{(2\pi)} &\sim
                      (k_{1}-k_{2})_{\mu}\hat{D}_{\rho}(q^{2})=
                      (k_{1}-k_{2})_{\mu}\left[D_{\rho}(q^{2}) + \beta D_{\rho'}(q^{2})\right].
                      \label{eq:amp2pi}
\end{align}
Here $q$ is the total momentum of the pionic pair, $k_{1,2}$ are the momenta of the final pions, the Lorentz structure of the expression is caused by the fact that these particles are in the $P$-wave state, while the propagator of the virtual resonance is written in the Flatte form \cite{Flatte:1976xu, Kuhn:1990ad,Kuhn:2006nw}
\begin{align}
  D_{\rho}(q^{2}) &= \frac{m_{\rho}^{2}}{m_{\rho}^{2}-q^{2}-i m_{\rho}\Gamma_{\rho}(q^{2})},
  \label{eq:Drho}
\end{align}
where the energy-dependent $\rho$-meson width
\begin{align}
  \Gamma_{\rho}(q^{2}) &= \left( \frac{1-4m_{\pi}^{2}/q^{2}}{1-4m_{\pi}^{2}/m_{\rho}^{2}}\right)^{3/2}\Gamma_{\rho}^{exp}
\end{align}
was introduced. The propagator of the excited $\rho$ is defined in the similar way, and, according to paper \cite{Schael:2005am}, the mixing parameter $\beta$ is equal to
\begin{align}
  \beta &= -0.108.
\end{align}
The normalization of the amplitude (\ref{eq:amp2pi}) is determined by the experimental value of the branching fraction
\begin{align}
  \Br(\tau\to\nu_{\tau}\pi^{-}\pi^{0}) &= 25.49\%.
\end{align}
Obtained in this way spectral function is shown in the right plane of Figure \ref{fig:rt2pi}.

\begin{figure}
\begin{centering}
\includegraphics[width=0.45\textwidth]{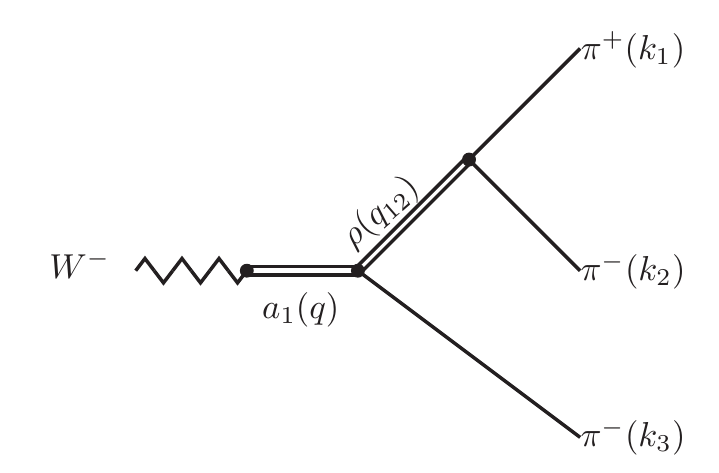}\includegraphics[width=0.45\textwidth]{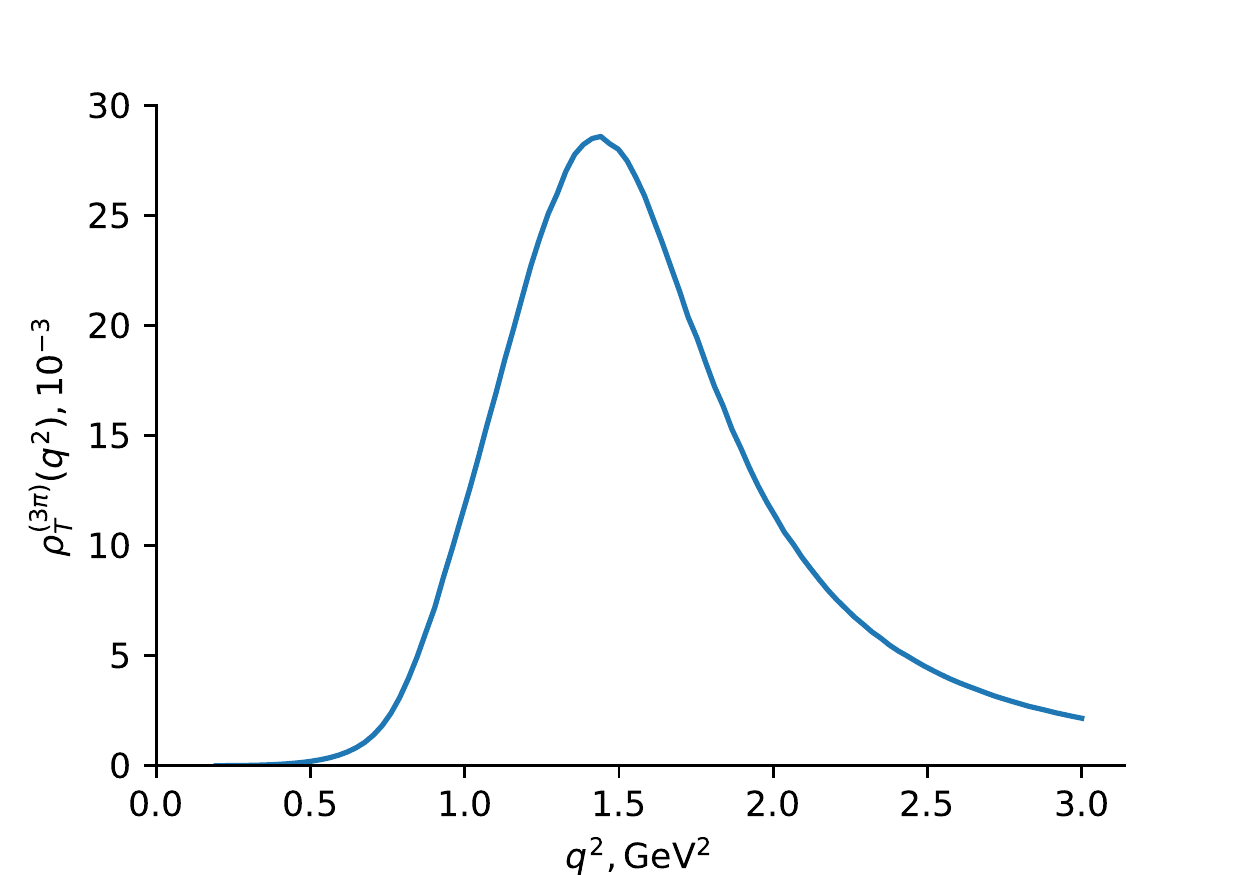}
\par\end{centering}
\caption{Feynman diagram for the  $W\to\pi^{-}\pi^{-}\pi^{+}$ transition and the corresponding spectral function\label{fig:rt3pi}}
\end{figure}

If production if three $\pi$ mesons is considered, in the resonance approximation one can take into account only diagram shown in figure \ref{fig:rt3pi}. The analytical expression for the corresponding amplitude is
\begin{align}
\A_{\mu}^{(3\pi)} & \sim D_{a_{1}}\left(q^{2}\right)\hat{D}_{\rho}\left[q_{12}^{2}\right]\left[g_{\mu\nu}-\frac{q_{\mu}q_{\nu}}{q^{2}}\right](k_{1}-k_{2})^{\nu}+\left\{ k_{2}\leftrightarrow k_{3}\right\} ,
\end{align}
where the particles' momenta are shown on the figure, symmetrization is performed over the identical $\pi^{-}$ mesons, the virtual $a_{1}$ meson propagator is equal to
\begin{align}
  D_{a_{1}}\left(q^{2}\right) & = \frac{m_{a_{1}}^{2}}{m_{a_{1}}^{2}-q^{2}+im_{a_{1}}\Gamma_{a_{1}}(q^{2})},\label{eq:Da1}
\end{align}
and the propagator of the $\rho$ meson and its excitations was introduced earlier in eq.~(\ref{eq:Drho}). The normalization branching fraction is equal to
\begin{align}
  \Br(\tau\to\nu_{\tau}\pi^{-}\pi^{-}\pi^+) &= 9.31\%.
\end{align}
and the $q^2$ dependence of the spectral function $\rho_T^{(3\pi)}$ can be found on the right panel of figure \ref{fig:rt3pi}.

\begin{figure}
\includegraphics[width=0.45\textwidth]{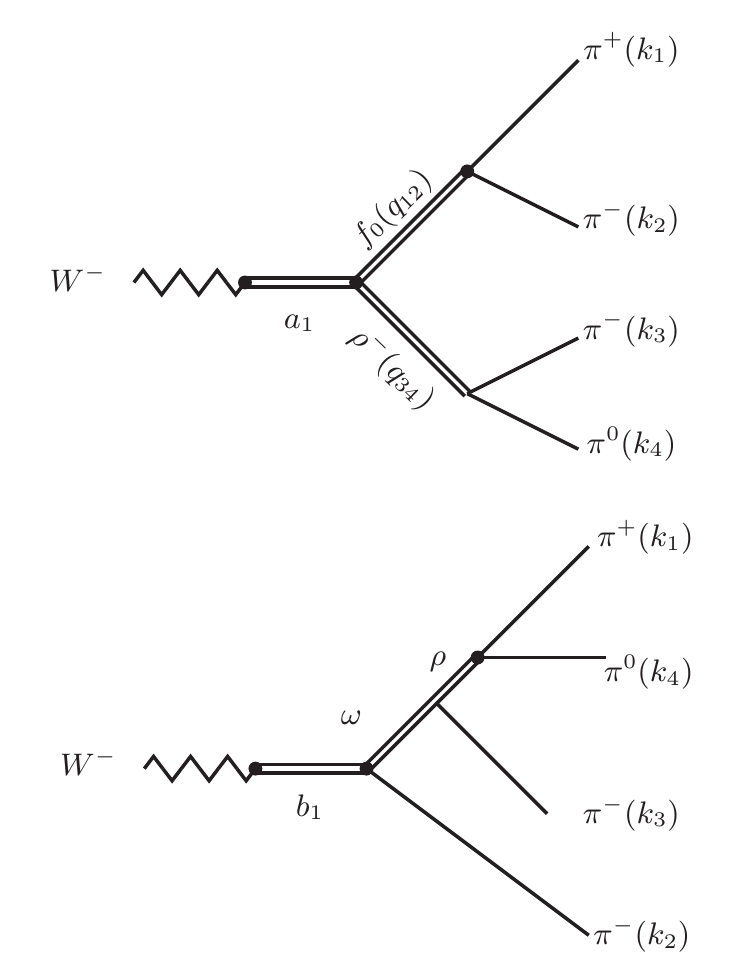}\includegraphics[width=0.45\textwidth]{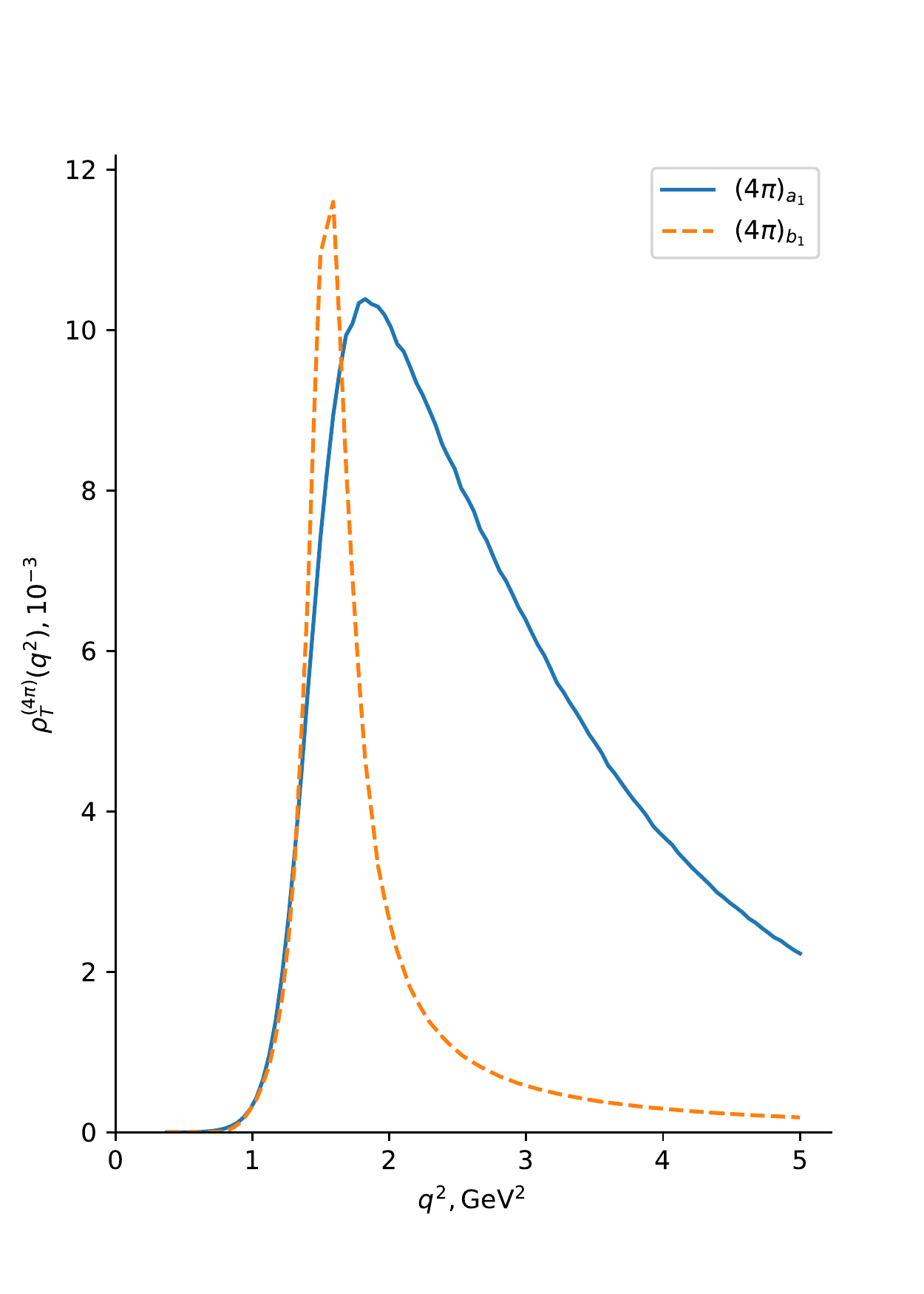}
\caption{Feynman diagram for the  $W\to\pi^{-}\pi^{-}\pi^{+}\pi^{0}$ transition and the corresponding spectral function $W\to4\pi$\label{fig:rt4pi}}
\end{figure}

According to Feynman diagrams shown in figure \ref{fig:rt4pi} production of $4\pi$ system can occur via either $a_1$ or $b_1$ virtual resonance. The matrix element of the first process can be written in the form
\begin{align}
\A_{\mu}^{(4\pi,a_{1})} & \sim  D_{a_{1}}(q^{2})D_{f}\left(q_{12}^{2}\right)D_{\rho}\left(q_{34}^{2}\right)\left[g_{\mu\nu}-\frac{q_{\mu}q_{\nu}}{q^{2}}\right]\left(k_{3}-k_{4}\right)^{\nu}+\left\{ k_{2}\leftrightarrow k_{3}\right\} ,\label{eq:amp4pia1}
\end{align}
where $q_{12} = k_1 + k_2$ and $q_{34} = k_3+k_4$ are the momenta of $f_0$ and $\rho$ mesons respectively and the Flatte parametrization of $f_0$ meson propagator is equal to
\begin{align}
D_{f}\left(q\right) & = \frac{m_{f}^{2}}{m_{f}^{2}-q^{2}+im_{f}\Gamma_{f}\left(q^{2}\right)},\qquad\Gamma_{f}\left(q^{2}\right)=\left(\frac{1-4m_{\pi}^{2}/q^{2}}{1-4m_{\pi}^{2}/m_{f}^{2}}\right)^{1/2}\Gamma_{f}^{exp},\label{eq:Df}
\end{align}
As you can see, the exponent in the expression for the running width of the $f_0$ meson is different from that of $\rho$ meson [see equation (\ref{eq:Drho})]. The reason that in the decay of $f_0$ meson the final particles are in the $S$ wave. The amplitude of $W\to b_1 \to 4\pi$ transition, on the other hand, can be written as
\begin{align}
\A_{\mu}^{(4\pi,b_{1})} & \sim D_{b_{1}}\left(q^{2}\right)D_{\omega}\left(q_{123}^{2}\right)D_{\rho}\left(q_{12}\right)e_{\mu\nu\alpha}q_{123}^{\nu}q_{12}^{\alpha}\left(k_{1}-k_{2}\right)^{\beta},\label{eq:amp4pib1}
\end{align}
where $q_{123}$ and $q_{12}$ are the momenta of the virtual $\omega$ and $\rho$ mesons respectively and all propagators are defined in relations (\ref{eq:Drho}), (\ref{eq:Da1}), (\ref{eq:Df}). Due to the smallness of $\omega$ meson's width these two channels do not interfere with each other and the coupling constants can be determined from the branching fractions of the corresponding $\tau$ lepton decays:
\begin{align}
\Br\left[\tau\to a_{1}\nu_{\tau}\to4\pi\right] & =  2.74\%,\\
\Br\left[\tau\to b_{1}\nu_{\tau}\to4\pi\right] & =  1.8\%.
\end{align}

\begin{figure}
\includegraphics[width=0.45\textwidth]{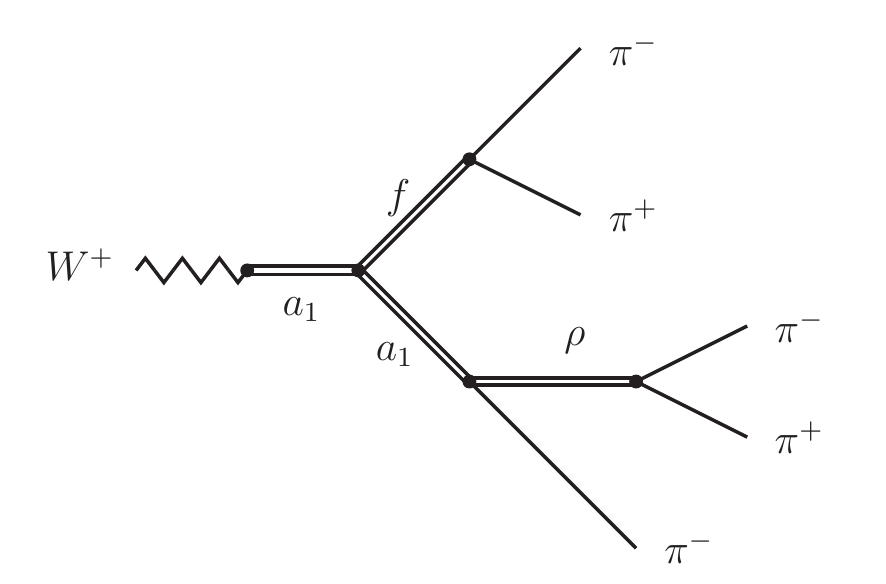}\includegraphics[width=0.45\textwidth]{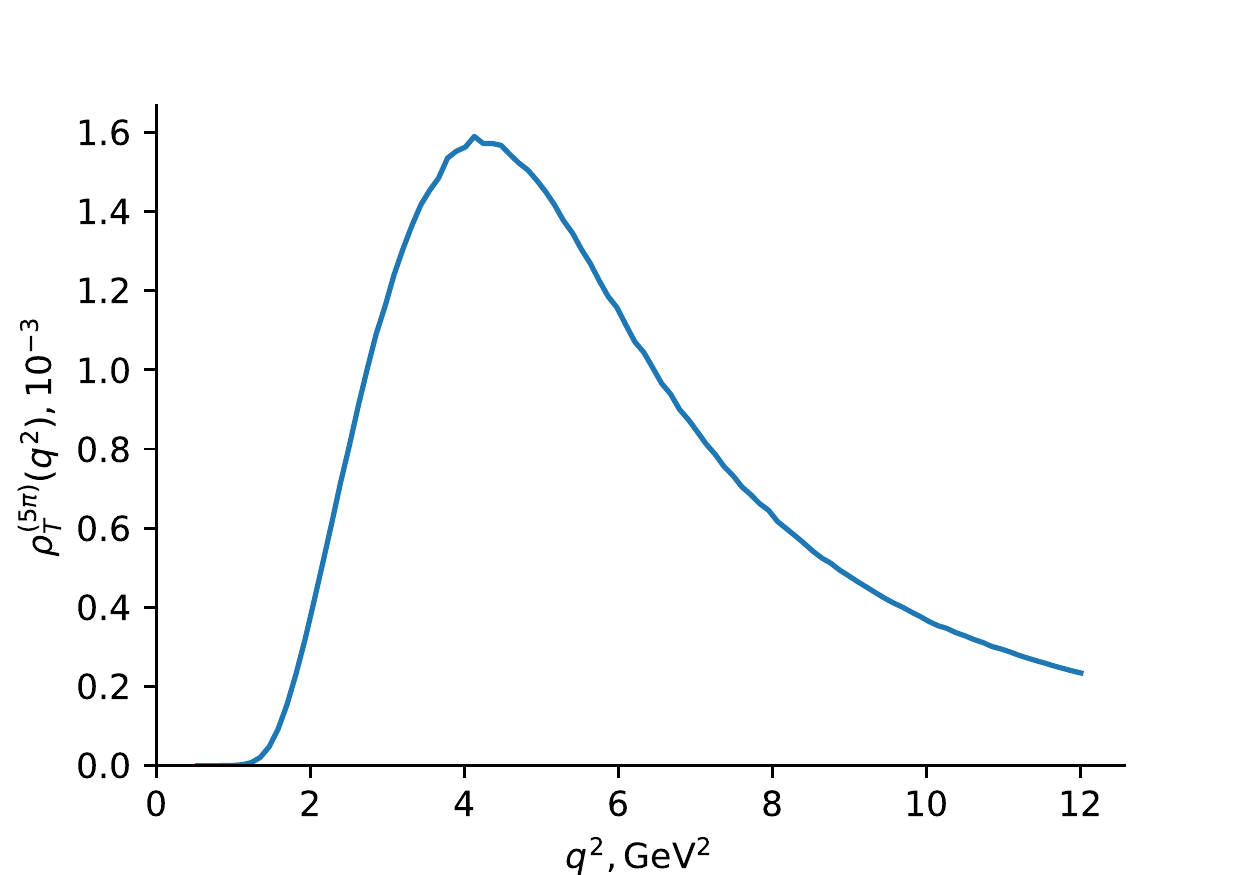}
\caption{Feynman diagram and spectral function of the  $W\to5\pi$ transition\label{fig:rt5pi}}
\end{figure}

Let us finally discuss production of the $5\pi$ meson system. The matrix element of this transition is (see the right pane of the figure \ref{fig:rt5pi} for the corresponding Feynman diagram)
\begin{align}
 \A_{\mu}^{(5\pi)} & \sim  D_{a_{1}}\left(q^{2}\right)D_{a_{1}}\left(q_{123}^{2}\right)D_{f}\left(q_{45}^{2}\right)D_{\rho}\left(q_{13}^{2}\right)\left[\frac{q_{\mu}q_{\nu}}{q^{2}}-g_{\mu\nu}\right]\left[\frac{q_{123}^{\nu}q_{123}^{\alpha}}{q_{123}^{2}}-g^{\nu\alpha}\right]\left(k_{1}-k_{3}\right)_{\alpha}+\text{permutation},
\end{align}
where the particles' momenta are shown on the figure and overall normalization can be determined from the branching fraction
\begin{align}
\Br\left(\tau\to\nu_{\tau}\pi^{+}\pi^{+}\pi^{-}\pi^{-}\pi^{-}\right) & = 8.27\times10^{-4}.
\end{align}
The spectral function itself is shown on the right panel of the figure.

With the help of presented above analytical expressions, spectral functions and form factors it is easy to obtain given in table \ref{tab:Br} branching fractions of the considered in our article decays. In these calculations the following values for $\Xi_{bc}^{+}$ mass and total lifetime \cite{Likhoded:1999yv, Berezhnoy:2018bde} were used:
\begin{align}
M_{\Xi_{bc}^{+}} & = 6.943\,\GeV,\qquad\tau_{\Xi_{bc}^{+}}=0.24\pm0.02\,\ps.
\end{align}
From the presented table it can be clearly seen the the branching fractions of some of the decays are rather large, so it could be possible to observe them experimentally. It is also worth mentioning that theoretical predictions depend strongly on the choice of the form factors' set. As a result new experimental data about these decays, could help us to see, which model describes better the real physics of the doubly heavy baryons.

\begin{table}
\begin{tabular}{c|ccc}
\hline
$\R$ & [On\_00] & [W\_17] &  [H\_20] \\
\hline
  $2\pi$ & $1.86$ & $0.41$ & $0.47$ \\ 
  $3\pi$ & $1.29$ & $0.29$ & $0.33$ \\ 
  $(4\pi)_{a_1}$ & $1.16$ & $0.27$ & $0.29$ \\ 
  $(4\pi)_{b_1}$ & $0.31$ & $0.07$ & $0.08$ \\ 
  $5\pi$ & $0.33$ & $0.07$ & $0.08$ \\
  \hline
\end{tabular}
\caption{The branching fractions of $\Xi_{bc}^{+}\to\Xi_{cc}^{++}\R$ decays (in \%)\label{tab:Br}}
\end{table}

\begin{figure}
\begin{centering}
\includegraphics[width=0.9\textwidth]{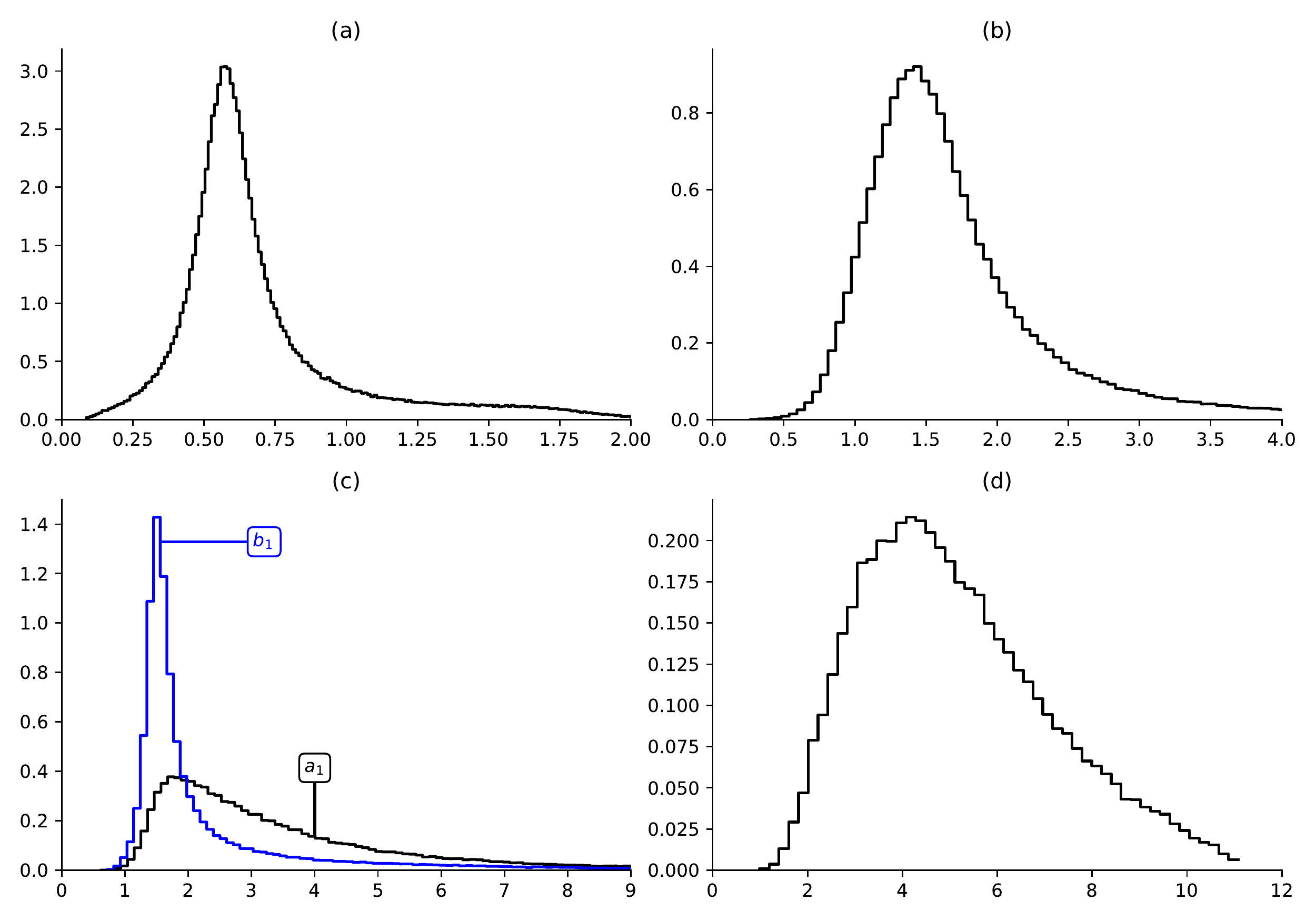}
\par\end{centering}
\caption{Normalized distributions of the $\Xi_{bc}^{+}\to\Xi_{cc}^{++}\R$ branching fractions over the squared transferred momentum $q^{2}$ (in $\GeV^{2}$) for different final states. Subplots (a), (b), (c), and (d) correspond to final states $\R=3\pi$, $3\pi$, $4\pi$, and $5\pi$ respectively. Only results of [On\_00] for factors set are given
\label{fig:q2_all}}
\end{figure}

Distributions over the squared mass of the light mesons' system are shown in figure \ref{fig:q2_all}. Our calculations show that the forms of these distributions only slightly depend on the choice of the form factors set, so on this figure only normalized result for [On\_00] FF set are shown.

\section{Distributions over the Other Kinematical Variables}
\label{sec:distr-over-other}

In this section we will discuss distributions over some other kinematical variables (such as the invariant mass of the pion pair). It is clear that these distributions require the complete information about the dynamics of the process, so the spectral function formalism cannot be used. Moreover, since we are working with the decays with high number of particles in the final state, one cannot use any analytical methods, only numerical calculations is appropriate. One of the convenient tools that can help in such situations is the Monte-Carlo generator EvtGen \cite{Ryd:2005zz,Lange:2001uf}, that is used by the LHCb collaboration. Our group has created the required software models and in the following we discuss the results obtained using these models. As it was mentioned above, the form of the normalized distributions depends only slightly on the choice of the form factors' set, so below we present only the result of \cite{Onishchenko:2000wf} parametrization.

\begin{figure}
  \centering
  \includegraphics[width=0.9\textwidth]{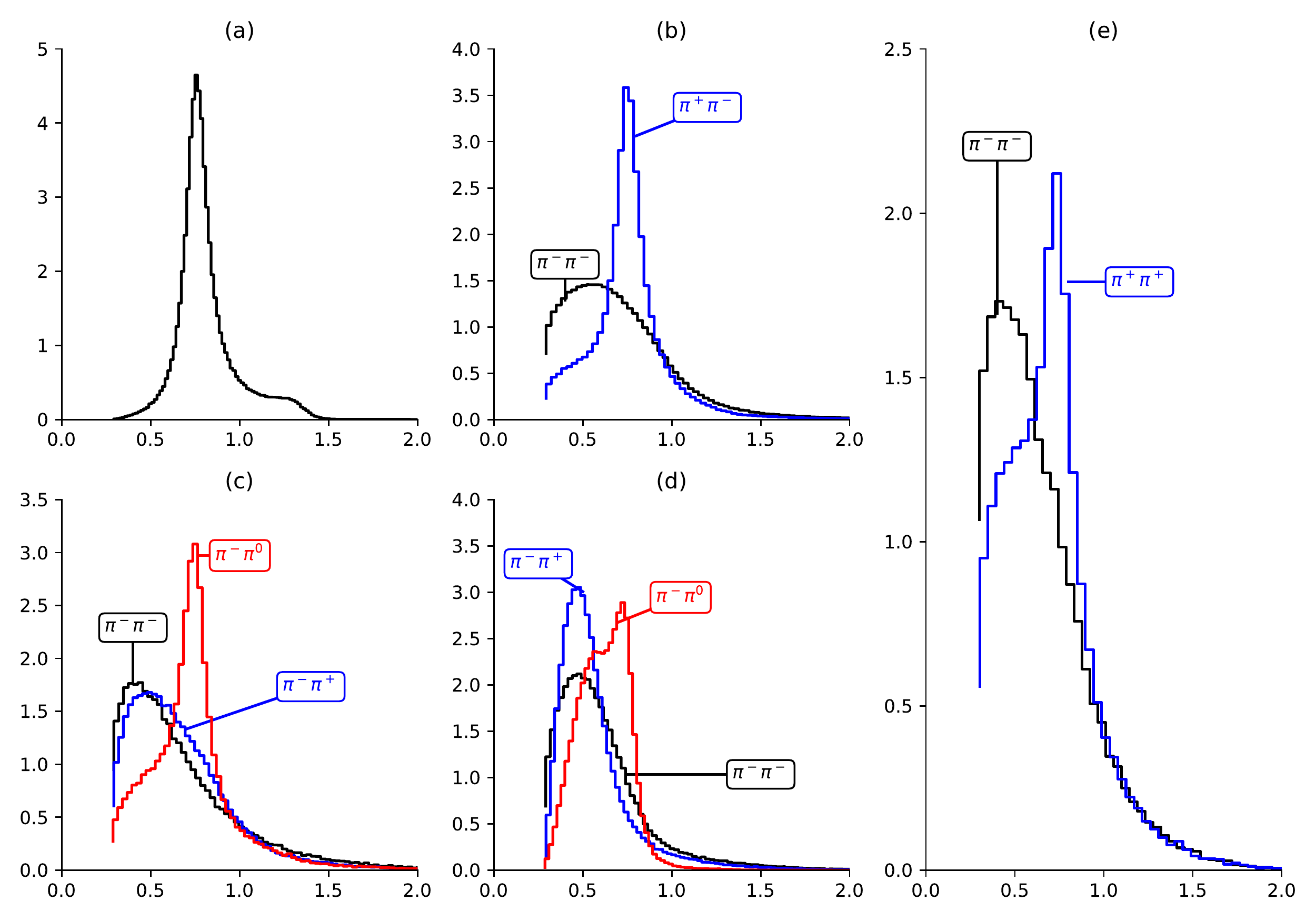}
  \caption{Normalized distributions of the decays $\Xi_{bc}^{+}\to\Xi_{cc}^{++}\R$ over the invariant mass  $\pi\pi$ system (in GeV). Subplots (a), (b), (c), (d), and (e) correspond to final states $\R=2\pi$, $3\pi$, $(4\pi)_{a_1}$, $(4\pi)_{b_1}$ and $5\pi$ respectively}
  \label{fig:m_2pi}
\end{figure}

Let us first consider distributions over the invariant mass of the $\pi$ mesons' pair (see figure \ref{fig:m_2pi}). It is clear from this figure that the form of the distribution depends both on the final state $\R$ and the charged of the mesons. In the case of three $\pi$ mesons production, for example, you can see a clear peak in $m_{2\pi}\approx m_\rho$ region in $m_{\pi^+\pi^-}$ distribution [see subplot (b) of this figure], while in the case of $\Xi_{bc}^{+}\to\Xi_{cc}^{++}a_1\to\Xi_{cc}^{++}+4\pi$ there is no sign of such a peak. The reason for such behavior is that in the former case $\pi^+\pi^-$ pair is produced in the decay of virtual $\rho$ meson [see the diagram shown in fig. \ref{fig:rt3pi}], while in the latter case this pair is produced in $f_0$ meson decay [see presented in fig. \ref{fig:rt4pi} diagram]. Since the width of the $f_0$ meson is rather large, the peak on the corresponding $m_{2\pi}$ distribution is hardly visible. According to the same diagram, on the other hand, $\pi^-\pi^0$ pair can be produced in $\rho^-$ decay and the corresponding peak is clearly seen in shown in fig.~\ref{fig:m_2pi}(c) distribution over the invariant mass of this pair. It is easy to check that the same behavior is observed for all other reactions and final states: whenever any $\pi$ meson pair is produced in the decay of $\rho$ meson, there is a peak (probably modified a little bit by the combinatorial background) in the corresponding distribution.

\begin{figure}
  \centering
  \includegraphics[width=0.7\textwidth]{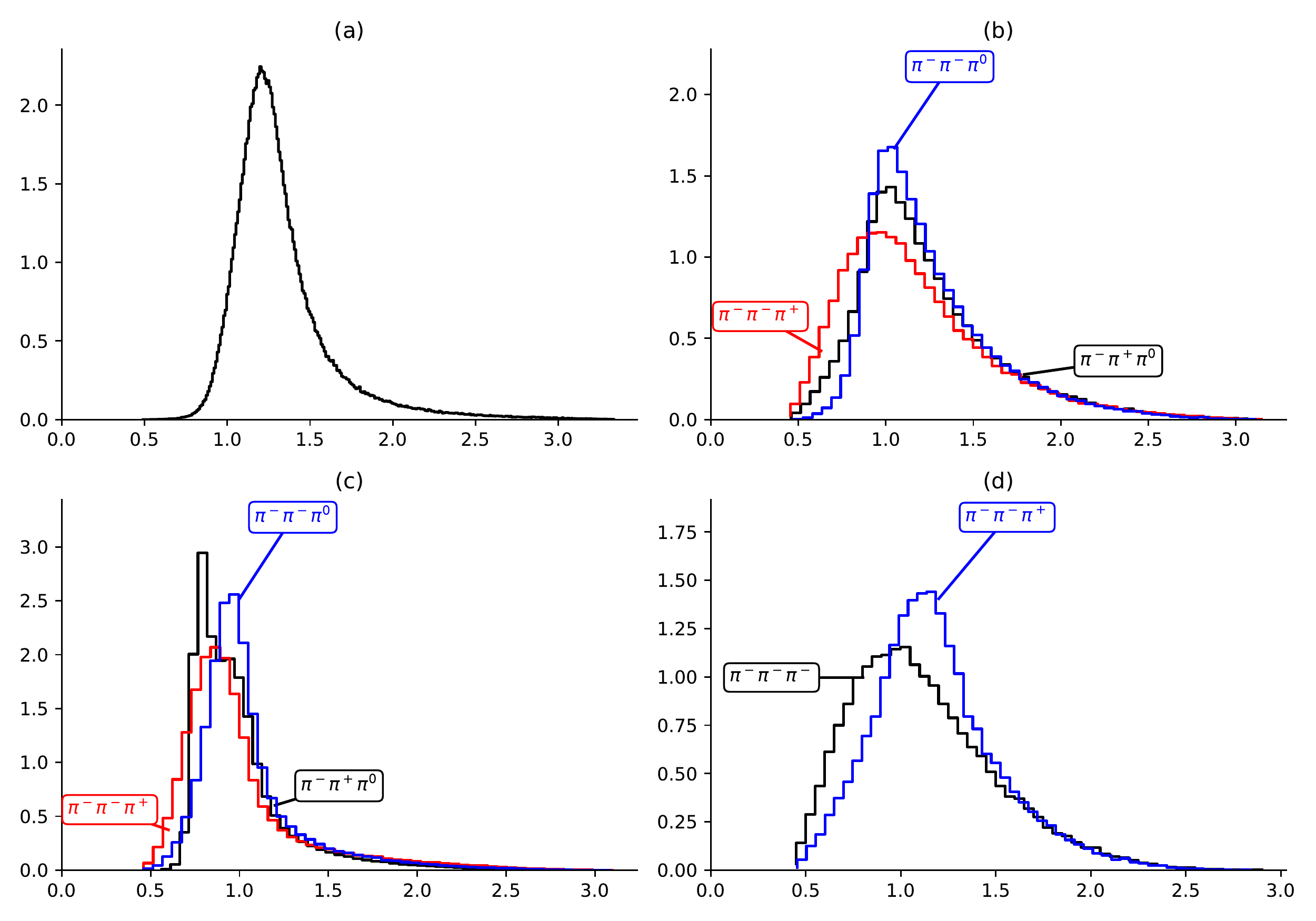}
  \caption{
    Normalized distributions of the reactions $\Xi_{bc}^+\to\Xi_{cc}^{++}\R$ over the invariant mass of three $\pi$ mesons (in GeV). Subplots (a), (b), (c), and (d) correspond to the final states $\R=3\pi$, $(4\pi)_{a_1}$, $(4\pi)_{b_1}$, and $5\pi$ respectively.}
  \label{fig:m_3pi}
\end{figure}

The same is true also for the distributions over the invariant masses of three $\pi$ mesons. In figure \ref{fig:m_3pi}(c), for example, we can see a peak in $m_{\pi^+\pi^-\pi^0}$ distribution, caused by shown in the diagram $\omega$ resonance (due to the combinatorial background the form of this peak is not symmetric). There is also a clear peak in $m_{\pi^-\pi^-\pi^+}$ distribution in the case of $\R=5\pi$ final state (figure \ref{fig:m_3pi}(d)], that corresponds to $a_1$ resonance in diagram \ref{fig:rt5pi}.

\begin{figure}
  \centering
  \includegraphics[width=0.7\textwidth]{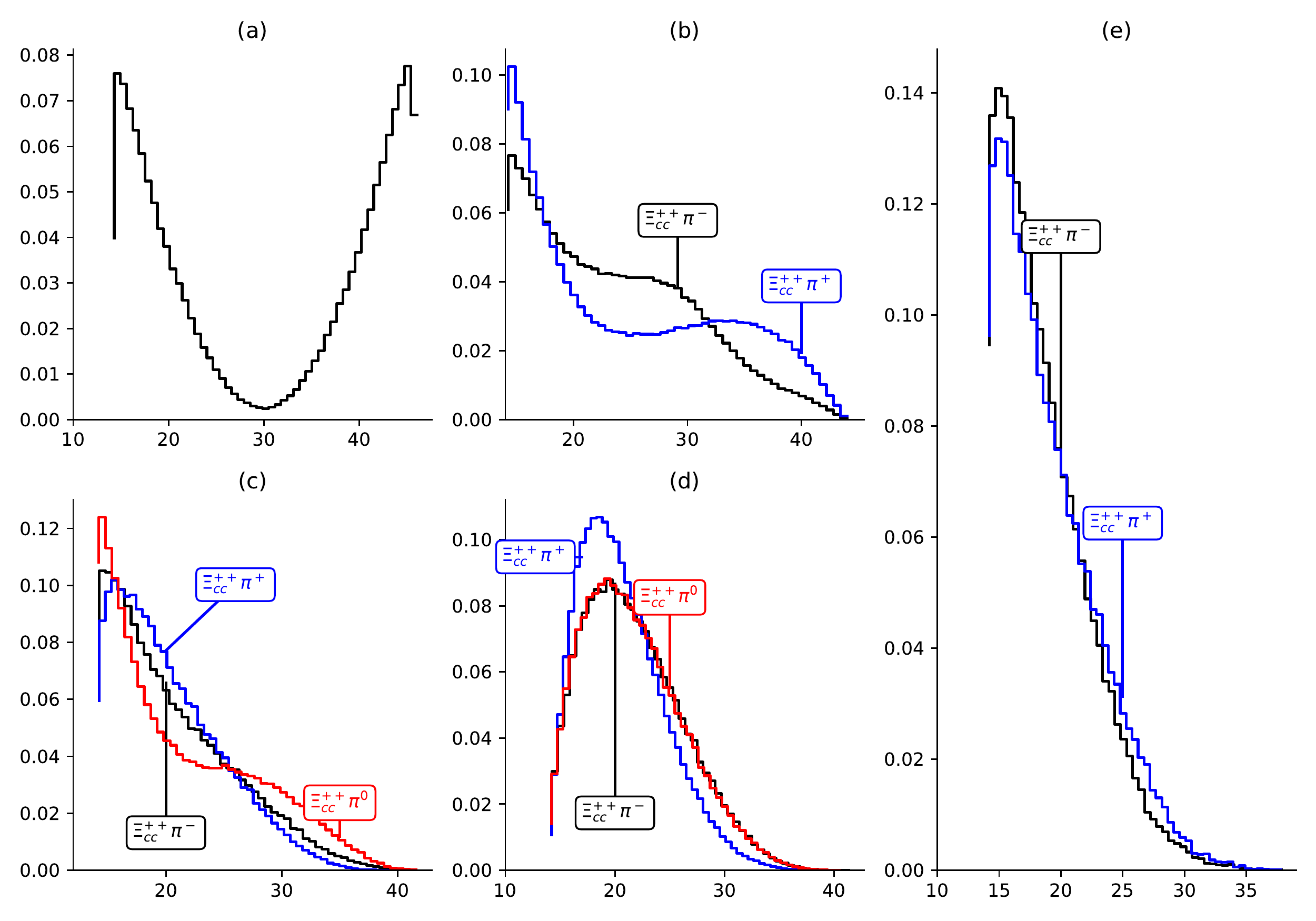}
  \caption{Normalized distributions of the decays $\Xi_{bc}^{+}\to\Xi_{cc}^{++}\R$ over the squared invariant mass  $\Xi_{cc}\pi$ system (in $\GeV^2$). Subplots (a), (b), (c), (d), and (e) correspond to final states $\R=2\pi$, $3\pi$, $(4\pi)_{a_1}$, $(4\pi)_{b_1}$ and $5\pi$ respectively}
  \label{fig:m2Bpi}
\end{figure}

Let us finally consider distributions over the invariant mass of $\Xi_{cc}\pi$ pair. These distributions are shown in figure \ref{fig:m2Bpi} and their behavior is much more interesting. In the case of $\R=\pi^+\pi^-$ final state, for example, you can see two clear peaks near ends of the allowed region. It is clear that this distribution should be symmetric (the reflection the plot corresponds to the interchange $\pi^+ \leftrightarrow \pi^0$ that does not change the matrix element \eqref{eq:amp2pi} ), but the reason for the peaks is not evident. You can see also, that in the case of $\pi^+\pi^-\pi^-$ final state there is some peak in $m_{\Xi\pi^-}$ distribution. According to presented in the previous section diagrams and matrix elements there are no virtual resonances in the corresponding channels, so we can say that the reason for such a behavior is some interplay of the hadronic matrix elements of $\Xi_{bc}\to\Xi_{cc}W$ and $W\to\R$ transitions, phase space region, etc.

% stopped here 

\section{Conclusion}
\label{sec:conclusion}

In the presented article we analyze some of the exclusive decays of $\Xi_{bc}^+$ baryon with production of light mesons. In our previous paper \cite{Gerasimov:2019jwp} we have considered this type of doubly heavy baryons' decays with the help of spectral function formalism and calculated their branching fractions and distributions over the invariant mass of the light mesons' system. According to presented in that article results the branching fractions of some of such decays are high enough to observe them experimentally. It is clear, however, that for analysis of the experimental data it is required to know  distributions over other kinematical variables. Such results cannot be obtained in the framework of the approach used in our previous paper, so a more detailed theoretical models are required.

In the presented paper we have preformed such an analysis concentrating on some exclusive decays of $\Xi_{bc}^{+}$. This particular baryon was chosen since one could expect the experimental observation of this particle in the nearest future. In our work we have calculated the branching fractions of the reactions $\Xi_{bc}^{+}\to\Xi_{cc}^{++}\R$, where light mesons' system $\R$ could be $2\pi$, $3\pi$, $4\pi$, and $5\pi$. For all these reactions the branching fractions were calculated and the distributions over different kinematical variables are presented. According to our results in the distributions over the masses of systems of two or three $\pi$ mesons some peaks caused by the virtual resonances (such as $\rho$, $\omega$, $a_1$, etc) should be clearly seen. It could be even more interesting to study distributions over $\Xi_{cc}\pi$ masses, for which our model predicts some additional peaks that do not correspond to any intermediate particles.

It is worth mentioned that in our work the factorization approach was used, in which the matrix elements of the considered processes are written as a product of $\Xi_{bc}\to\Xi_{cc}W$ and $W\to\R$ transitions. This assumption looks absolutely suitable for the similar decays of $B_c$ meson. When baryons' decays are discussed, however, the non-factorizable diagrams should give some contributions. Although these contributions are color-suppressed, their effect could be noticeable. In our future work we are planing to consider these corrections in more details.

The authors would like to thank A.V. Berezhnoy for useful discussions. This research was done with support of RFBR grant № 20-02-00154 A.

% \bibliographystyle{apsrev4-1}
% \bibliography{litr_xibc}


\end{document}